\documentclass[12pt]{article}

\usepackage[margin=1in]{geometry}

\usepackage{setspace}

\usepackage{graphicx}

\usepackage{enumitem}

\usepackage{microtype}
\usepackage{subfigure}
\usepackage{booktabs}
\usepackage{comment}
\usepackage{wrapfig}
\usepackage{arydshln}
\usepackage{enumitem}
\usepackage{multirow}
\usepackage{url}
\usepackage{natbib}
\usepackage{authblk}

\RequirePackage[colorlinks,citecolor=blue,linkcolor=blue,urlcolor=blue,pagebackref]{hyperref}

\usepackage{tabularx}
\usepackage{amsmath}
\usepackage{amssymb}
\usepackage{mathtools}
\usepackage{amsfonts}
\usepackage{bm}
\usepackage{bbm}

\allowdisplaybreaks

\usepackage{algorithm}
\usepackage{algorithmic}

\def\1{\bm{1}}

\DeclareMathAlphabet{\mathsfit}{\encodingdefault}{\sfdefault}{m}{sl}
\SetMathAlphabet{\mathsfit}{bold}{\encodingdefault}{\sfdefault}{bx}{n}

\usepackage{amsthm}

\theoremstyle{plain}

\usepackage[textsize=tiny]{todonotes}
\usepackage{multirow}
\usepackage{wrapfig}
\usepackage{subfigure}

\usepackage{xcolor}
\newcount\Comments  
\newcommand{\kibitz}[2]{\ifnum\Comments=1\textcolor{#1}{#2}\fi}

\usepackage{listings}

\lstdefinestyle{algospecformat}{
  basicstyle=\ttfamily\scriptsize,
  breaklines=true,
  columns=fullflexible,
  frame=single,
  keepspaces=true,
  showstringspaces=false,
  xleftmargin=0.5em,
  xrightmargin=0.5em
}

\usepackage{placeins}

\usepackage[capitalize,noabbrev]{cleveref}

\allowdisplaybreaks

\title{Which Algorithm Specification Formats Help Language Models Implement Machine Learning Algorithms?}
\author{Masahiro Kato\thanks{Email: \texttt{mkato-csecon@g.ecc.u-tokyo.ac.jp}}$\,$}
\author{Taka Kato\thanks{Email: \texttt{taka@np-hard.co.jp}}$\,$}

\affil{${}^*$Data Analytics Department, Mizuho-DL Financial Technology, Co., Ltd.}
\affil{${}^\dagger$NP-hard Inc.}

\date{\today}

\begin{document}

\maketitle 

\begin{abstract}
Large language models (LLMs) are increasingly used to implement algorithms from research manuscripts, but papers often leave implementation choices implicit. This study examines how the written format of an algorithm specification affects first-pass LLM implementation accuracy. We compare ordinary prose, LaTeX algorithm-style pseudocode, PDF-like extracted pseudocode, Markdown fields, YAML-like specifications, JSON-like specifications, and Python code stubs across five machine learning tasks, three models, and four experimental settings, yielding 4,020 generated implementations. Hidden tests evaluate details that often determine correctness, including tie-breaking, array shapes, numerical rules, return structures, and invalid-input behavior. Under the core-information setting, LaTeX algorithm-style pseudocode has the largest average format effect, with YAML-like specifications and ordinary prose close behind. Under complete information, GPT-5.4 mini shows no format differences in the matched comparisons, whereas Gemma 3 4B and Llama 3.2 3B still do. Code stubs do not consistently improve correctness despite specifying the function signature. The results support a writing recommendation: authors should state the interface, computation steps, numerical rules, and boundary-case behavior explicitly, instead of relying on a particular surface format to carry those details.
\end{abstract}

\section{Introduction}
Large language models (LLMs) are increasingly used in workflows that extract implementation details from research manuscripts, generate code, run tests, and revise failed attempts. This study investigates how algorithm specifications in academic articles should be written when they may be used as inputs to LLM-based implementation. We test whether conventional pseudocode is sufficient for this purpose, or whether other representation formats convey implementation-relevant details more effectively.

This motivation is connected to recent benchmarks on paper-to-code generation and research reproduction. PaperBench evaluates agents on the replication of $20$ ICML 2024 Spotlight and Oral papers and decomposes the replications into 8,316 gradable subtasks \citep{Starace2025paperbench}. ResearchCodeBench builds $212$ coding challenges from academic articles and reports that the best evaluated models solve fewer than $40$ percent of the tasks \citep{Hua2025researchcodebench}. SciReplicate-Bench evaluates code generation from algorithm descriptions in recent natural language processing papers and emphasizes algorithm comprehension and coding expertise as separate requirements \citep{Xiang2025scireplicatebench}. \citet{Movassaghi2026fromarticles} asks whether publications alone can serve as specifications for regenerating core algorithms, and finds that focused, well-described methods can often be reproduced reliably. Together, these studies point to a common source of error: implementation details that are missing, scattered, or underspecified in the paper. This study isolates one part of that problem by changing the representation format of the algorithm specification while holding the target function and tests fixed. 

\subsection{Contributions}
This study asks what authors should specify when an algorithm description may be used for LLM-based implementation. We compare formats within matched model, task, information-setting, prompt, and repeat groups, so each comparison changes the written form while keeping the target implementation and tests fixed. The results show that format can affect hidden-test correctness, but the direction of the effect depends on the model and on how much implementation information the prompt contains.

We evaluate seven representation formats: ordinary prose, LaTeX algorithm-style pseudocode, PDF-like extracted pseudocode, Markdown fields, YAML-like specifications, JSON-like specifications, and Python code stubs. The main comparisons use two information settings. The core-information setting gives the function name, Python signature, purpose, inputs, outputs, and computation steps. The complete-information setting adds configuration values, numerical rules, and invalid-input behavior.

Our main finding is conditional rather than a single ranking. With core information, LaTeX algorithm-style pseudocode has the largest average format effect, while YAML-like specifications and ordinary prose remain close. With complete information, GPT-5.4 mini shows no format differences in the matched comparisons, but Gemma 3 4B and Llama 3.2 3B still do. Code stubs state the signature, but their empty bodies do not supply the computation rules needed for hidden-test correctness. The resulting recommendation is to state implementation details explicitly, not to standardize on one syntax.

In summary, this study makes four contributions:
\begin{enumerate}
\item We define a paired comparison design that varies only the representation format within each comparison, holding everything else about the generation fixed.
\item We report results from 4,020 generated implementations across five machine learning implementation tasks, seven representation formats, three models, and four experimental settings.
\item We show that the ordering of formats by score depends on the model and on the amount of implementation information in the prompt. The results separate interface explicitness from implementation correctness: a code stub can state the function boundary without giving enough computational detail to pass hidden tests.
\item We examine whether expected-output examples, executable public tests, and the removal of specific information fields change implementation accuracy. These analyses show that additional information does not have a uniform effect in a single generation without repair.
\end{enumerate}
\subsection{Related Work}
There are several lines of related work, including paper-to-code and research reproduction, code-generation evaluation, and prompt format and specification structure.

\paragraph{Paper-to-code and research reproduction.}
PaperBench, ResearchCodeBench, and SciReplicate-Bench evaluate systems that use papers as inputs for code generation or research reproduction \citep{Starace2025paperbench,Hua2025researchcodebench,Xiang2025scireplicatebench}. These benchmarks report substantial failures when implementation details are missing or hard to recover. Movassaghi et al. provide a complementary positive result: publications can be sufficient for many focused, well-described core algorithms \citep{Movassaghi2026fromarticles}. This study varies the specification format directly, rather than ranking agents or asking whether an entire paper is enough.

\paragraph{Code-generation evaluation.}
HumanEval introduced function-level evaluation in which generated programs are judged by tests and summarized with pass rates \citep{Chen2021evaluatinglarge}. MBPP evaluates short Python programming tasks from natural-language descriptions \citep{Austin2021programsynthesis}. SWE-bench extends code-generation evaluation to real software issues and repository-level tests \citep{Jimenez2024swebench}. This study follows the same functional-correctness principle and evaluates implementations by public and hidden tests rather than by text similarity.

\paragraph{Prompt format and specification structure.}
Prompt formatting can change model behavior across tasks and models, including code-related tasks \citep{He2024doesprompt}. However, prompt-format studies do not by themselves answer which paper-level algorithm specification format should be used for implementation. This study therefore evaluates representation formats in an implementation experiment with tests that are not shown to the model.
\section{Research Questions}
\label{sec:research_questions}

We ask how the written form of an algorithm specification affects LLM-generated implementations. The object of comparison is the representation format, not the models themselves. The experiment addresses five questions:
\begin{description}[leftmargin=1.7em]
\item[RQ1.] Does the written form of the same algorithm specification affect hidden-test correctness when the prompt contains the function name, Python signature, purpose, inputs, outputs, and computation steps?
\item[RQ2.] Do format differences remain when the prompt also contains configuration values, numerical rules, and invalid-input behavior?
\item[RQ3.] Does the ordering of formats by score differ across GPT-5.4 mini, Gemma 3 4B, and Llama 3.2 3B?
\item[RQ4.] Do expected-output examples and executable public tests improve implementation accuracy in a single generation without repair?
\item[RQ5.] For YAML-like specifications, JSON-like specifications, and Python code stubs, how does removing shapes, configuration values, numerical rules, or invalid-input behavior change implementation accuracy?
\end{description}

RQ1--RQ3 concern the main comparison of representation formats. RQ4 asks whether adding examples or public tests helps the first generated implementation. RQ5 examines whether specific information fields affect generated code in structured specifications.

\section{Experimental Design}
\label{sec:experimental_design}

This section describes the tasks, representation formats, prompt information, models, evaluation measures, and experimental settings. The design keeps the target function and tests fixed while changing the representation format shown to the model.

\subsection{Tasks and Hidden Tests}
\label{sec:tasks_tests}
We use five function-level machine learning tasks. The tasks are intentionally small. The goal is to test whether representation formats transmit implementation details; building large machine learning systems is out of scope here. Each task has a compact function signature, but correctness still depends on choices about ties, invalid inputs, empty inputs, array shapes, return types, and numerical computation. Table~\ref{tab:tasks} lists the tasks used in the main format comparisons.

\begin{table}[htbp]
\centering
\caption{Implementation tasks used in the main comparison. Hidden tests are not shown to the model and are used only for evaluation.}
\begin{tabularx}{\linewidth}{p{0.25\linewidth}p{0.30\linewidth}X}
\toprule
Task & Required function & Main checks in hidden tests \\
\midrule
Top-k distribution & \texttt{top\_k\_distribution} & Top-\(k\) selection, smaller-index tie-breaking, temperature, stable softmax, invalid inputs, empty logits, and input preservation \\
Classification metrics & \texttt{classification\_metrics} & Accuracy, macro precision, macro recall, macro F1, zero division, absent labels, and length mismatch \\
K-means & \texttt{kmeans} & Squared-distance assignment, smaller-index tie-breaking, empty clusters, convergence, invalid inputs, and preservation of initial centers \\
Scaled dot-product attention & \texttt{scaled\_dot\_product\_attention} & Score computation, softmax over the key dimension, mask behavior, broadcasting, numerical stability, and shape mismatch \\
Beam search selection & \texttt{beam\_search\_select} & Beam expansion, score ordering, lexicographic tie-breaking, return type, beam width larger than the candidate count, and invalid inputs \\
\bottomrule
\end{tabularx}
\label{tab:tasks}
\end{table}

For each task, we prepare public tests and hidden tests. Public tests are short tests that may be shown to the model in the public-tests setting. Hidden tests are evaluation tests that are never shown to the model. They are used only after the model has generated an implementation. A hidden test checks whether the generated function follows details that are easy to omit from prose or pseudocode, such as tie-breaking rules, invalid-input behavior, input preservation, softmax axes, and exact return structures.

In \texttt{top\_k\_distribution}, the hidden tests check whether the function keeps exactly \(\min(k,\mathrm{len}(\texttt{logits}))\) entries and assigns probability zero to the other entries. They also check the tie-breaking rule: when logits are equal, the smaller index must be retained first. The tests verify that temperature is applied before softmax, that softmax is computed stably by subtracting the maximum retained logit, and that the probabilities sum to one. They also check boundary cases, including \(k\) larger than the number of logits, empty logits, non-one-dimensional logits, \(k < 1\), and nonpositive temperature. One hidden test also checks that the input array is not modified.

In \texttt{classification\_metrics}, the hidden tests check whether the function returns the four required entries: \texttt{accuracy}, \texttt{macro\_precision}, \texttt{macro\_recall}, and \texttt{macro\_f1}. They verify that macro averages are computed over the labels supplied by the user, rather than only over labels appearing in the data. They also check the zero-division rule: if a label has no predicted positives, its precision contribution is defined as \(0.0\); if it has no true instances, its recall contribution is defined as \(0.0\); and if both precision and recall are zero, its F1 contribution is defined as \(0.0\). The tests include labels that are absent from both \texttt{y\_true} and \texttt{y\_pred}, and they check that a length mismatch between \texttt{y\_true} and \texttt{y\_pred} raises \texttt{ValueError}.

In \texttt{kmeans}, the hidden tests check deterministic assignment and update rules. The implementation must use squared Euclidean distance for assigning points to centers, and when a point is equally close to two centers, the smaller cluster index must be chosen. If a cluster receives no points, its previous center must be kept rather than replaced by zeros or by another point. The tests also check convergence behavior, the shape of the returned label and center arrays, and whether the initial centers are left unchanged. Invalid cases include an inconsistent shape for \texttt{initial\_centers}, \(k < 1\), \(\texttt{max\_iter} < 1\), negative tolerance, and an empty data matrix.

In \texttt{scaled\_dot\_product\_attention}, the hidden tests check the exact computation of attention scores and outputs. The score tensor must be computed as \(QK^\top/\sqrt{d}\), where the transpose is taken over the last two axes of \(K\), and \(d\) is the last dimension of \(Q\) and \(K\). The softmax must be applied over the key dimension, which is the last axis of the score matrix. The tests also check that a mask value of zero excludes the corresponding key before softmax, and that nonzero mask values include it. The mask must be broadcastable to the score shape. Additional tests use large scores to check numerical stability and shape mismatches to check that the function raises \texttt{ValueError}. The function must return only the attention output, not the attention weights.

In \texttt{beam\_search\_select}, the hidden tests check whether each existing beam is expanded by every vocabulary token and whether new scores are computed by adding token log probabilities to the old beam score. The selected beams must be sorted by descending score. If scores are tied, the beam with lexicographically smaller token sequence must come first. The tests also check that \texttt{beam\_width} may exceed the number of candidates, in which case all candidates are returned. The returned token sequence must be a Python list, not a tuple or a NumPy array. Invalid cases include empty beams, a nonpositive \texttt{beam\_width}, and a mismatch between the number of beams and the first dimension of \texttt{token\_log\_probs}.

\subsection{Representation Formats}
\label{sec:representation_formats}

A representation format is the written form used to present the same target function to the model. The target function, public tests, and hidden tests do not change across formats. We compare the following seven formats.

\begin{description}[leftmargin=1.8em]
  \item[Ordinary prose.] A plain English paragraph that describes the target function and its rules without named fields and without a code-block function definition.
  \item[LaTeX algorithm-style pseudocode.] A stepwise pseudocode block written in the style of algorithm environments commonly used in machine learning papers.
  \item[PDF-like extracted pseudocode.] A line-broken version of the pseudocode block, intended to approximate text copied or extracted from a paper PDF.
  \item[Markdown fields.] A document divided by Markdown headings and bullet lists, such as Function, Inputs, Outputs, and Steps.
  \item[YAML-like specification.] A field-based specification using YAML-style keys and indentation. The aim is to make field names explicit, not to require a complete formal schema.
  \item[JSON-like specification.] A field-based specification using JSON-style keys, strings, lists, and objects. It states the same fields in a JSON-like syntax.
  \item[Python code stub.] A Python function header with a docstring and an empty body. It gives the requested function signature but does not include the implementation.
\end{description}

\subsection{Short Example of the Seven Formats}
\label{sec:format_examples}

The following examples show shortened renderings of the same \texttt{top\_k\_distribution} task. Each example describes the same required function: keep the top-\(k\) logits, break ties by the smaller index, apply a stable softmax to the retained entries, set all other probabilities to zero, and raise \texttt{ValueError} for the specified invalid inputs. The experiment uses the full renderings generated from the canonical task specification. The examples below are shortened only to define the formats.

\paragraph{Ordinary prose.}
Ordinary prose is a single natural-language paragraph. It gives the function and its rules in continuous text, without field names, line numbers, or a Python function header.
\begin{lstlisting}[style=algospecformat]
Implement top_k_distribution(logits, k, temperature=1.0). The input logits is a one-dimensional NumPy array. Keep min(k, len(logits)) entries with the largest logits; ties are broken by the smaller index. Divide retained logits by temperature and apply a stable softmax only to those entries. Return a NumPy array with the same length as logits. Non-retained entries have probability 0. Raise ValueError for empty logits, non-one-dimensional logits, k < 1, or temperature <= 0.
\end{lstlisting}

\paragraph{LaTeX algorithm-style pseudocode.}
LaTeX algorithm-style pseudocode is a numbered sequence of steps written in the style of a LaTeX algorithmic block. It separates required inputs, the required output, and the computation steps, but it is not executable code.
\begin{lstlisting}[style=algospecformat]
\begin{algorithmic}[1]
\Require logits, k, temperature
\Ensure probability vector p with the same length as logits
\State Validate that logits is nonempty and one-dimensional.
\State Validate that k >= 1 and temperature > 0.
\State Set m = min(k, length(logits)).
\State Select m indices by descending logit value and smaller index on ties.
\State Apply stable softmax to logits[indices] / temperature.
\State Set p[i] = 0 for all non-selected indices and return p.
\end{algorithmic}
\end{lstlisting}

\paragraph{PDF-like extracted pseudocode.}
PDF-like extracted pseudocode is plain text that resembles an algorithm block after it has been copied or extracted from a PDF. It keeps the algorithm title, the input-output line, and numbered steps, but it does not keep the original LaTeX environment.
\begin{lstlisting}[style=algospecformat]
Algorithm 1 TopKDistribution
Input: logits, k, temperature. Output: p.
1. Validate logits, k, and temperature.
2. Let m be min(k, length(logits)).
3. Choose m entries by descending logit value; use smaller index for ties.
4. Compute stable softmax on the chosen logits divided by temperature.
5. Put the probabilities in their original positions and put 0 elsewhere.
6. Return p.
\end{lstlisting}

\paragraph{Markdown fields.}
Markdown fields use Markdown headings and bullet lists to separate parts of the specification. In this format, the function, inputs, and steps are written as named sections rather than as one continuous paragraph.
\begin{lstlisting}[style=algospecformat]
### Function
`top_k_distribution(logits, k, temperature=1.0)`

### Inputs
- `logits`: nonempty one-dimensional NumPy array
- `k`: positive integer
- `temperature`: positive float

### Steps
- Keep the top `k` logits, using the smaller index on ties.
- Apply stable softmax to retained logits divided by temperature.
- Return a full probability vector and set non-retained entries to 0.

### Errors
- Raise `ValueError` for empty or non-one-dimensional `logits`, `k < 1`, or `temperature <= 0`.
\end{lstlisting}

\paragraph{YAML-like specification.}
A YAML-like specification writes the same information with keys, indentation, and lists. The format makes the function name, signature, inputs, steps, output, and errors visible as separate fields.
\begin{lstlisting}[style=algospecformat]
function: top_k_distribution
signature: "def top_k_distribution(logits, k, temperature=1.0):"
inputs:
  logits: nonempty one-dimensional NumPy array
  k: positive integer
  temperature: positive float
steps:
  - keep min(k, len(logits)) entries
  - break ties by smaller index
  - apply stable softmax to retained logits divided by temperature
output: probability vector with zero probability outside the retained entries
errors: raise ValueError for invalid logits, k, or temperature
\end{lstlisting}

\paragraph{JSON-like specification.}
A JSON-like specification writes the information with braces, keys, strings, and arrays. It uses JSON-style syntax to state the function name, signature, inputs, rules, algorithm, and errors.
\begin{lstlisting}[style=algospecformat]
{
  "function": "top_k_distribution",
  "signature": "def top_k_distribution(logits, k, temperature=1.0):",
  "inputs": {
    "logits": "nonempty one-dimensional NumPy array",
    "k": "positive integer",
    "temperature": "positive float"
  },
  "rules": ["smaller index breaks ties", "non-retained entries are zero"],
  "algorithm": ["select top k entries", "apply stable softmax", "return full vector"],
  "errors": "raise ValueError for invalid logits, k, or temperature"
}
\end{lstlisting}

\paragraph{Python code stub.}
A Python code stub is a Python function declaration with a docstring and an empty body. It gives the exact function signature and a short description, but it does not contain the implementation.
\begin{lstlisting}[style=algospecformat]
def top_k_distribution(logits, k, temperature=1.0):
    """Keep top-k logits, break ties by smaller index,
    apply stable softmax, and set all other entries to 0.
    Raise ValueError for invalid logits, k, or temperature."""
    pass
\end{lstlisting}

\subsection{Prompt Information Settings}
\label{sec:prompt_information_settings}
The experiments vary both the representation format and the amount of information included in the prompt. The first main setting contains the function name, Python signature, purpose, inputs, outputs, and computation steps; we call it \emph{core information}. The second setting adds input and output shapes, configuration values, numerical rules, and invalid-input behavior; we call it \emph{complete information}.

Two additional settings start from complete information. The examples setting adds small input-output examples. The public-tests setting adds executable tests. In the field-removal experiment, we again start from complete information and remove one type of field at a time: shapes, configuration values, numerical rules, or invalid-input behavior.

\subsection{Models and Prompts}
\label{sec:models_prompts}

The experiment uses GPT-5.4 mini through the OpenAI API and two models served by Ollama: Gemma 3 4B and Llama 3.2 3B. The goal is to test whether changing the written format has similar effects across models of different sizes.

Each run used the same wrapper prompt. The wrapper required a single Python function, allowed only the Python standard library and NumPy, and specified the required error and return behavior. It did not add a function signature outside the rendered specification, so the signature had to come from the format-specific rendering.

\subsection{Evaluation Measures}
\label{sec:evaluation_measures}

Each generated implementation is evaluated by public tests and hidden tests. Public tests are shown only in the public-tests setting. Hidden tests are never shown to the model.

The primary score is the fraction of hidden tests passed. We call this score the hidden-test fraction. We also report a strict score that equals one only when all hidden tests pass. Other recorded quantities include whether all public tests pass, whether public tests pass but hidden tests fail, whether the requested function is defined, runtime errors, timeouts, failed test names, prompt length, and output length.

\subsection{Experimental Settings}
\label{sec:experimental_settings}
The experiment consists of four settings and 4,020 generated implementations in total. All settings use 10 repeats. Table~\ref{tab:scenarios} lists the purpose and size of each setting.

\begin{table}[htbp]
\centering
\caption{Overview of the four experimental settings. Each setting uses 10 repeats.}
\label{tab:scenarios}
\begin{tabular}{lr}
\toprule
Experiment & Implementations \\
\midrule
Core-information format comparison & 1,050 \\
Complete-information format comparison & 1,050 \\
Examples and public-test comparison & 840 \\
Field-removal comparison & 1,080 \\
\bottomrule
\end{tabular}
\end{table}

The first two experiments compare seven formats under core and complete information. The third adds examples or public tests for k-means and attention. The fourth removes one field family at a time from YAML-like specifications, JSON-like specifications, and Python code stubs.

\section{Results and Interpretation}
\label{sec:results}

This section reports the results from the four experiments. The analysis uses the hidden-test fraction defined in Section~\ref{sec:evaluation_measures} as the main score. We first define the comparison unit and summary measures. We then report the two main format comparisons, followed by model-specific results, the examples and public-tests experiment, the field-removal experiment, and two diagnostic checks.

\subsection{Comparison Unit and Summary Measures}
\label{sec:comparison_unit}

Because the goal is to compare representation formats, we compare formats within the same fixed comparison. A fixed comparison holds the model, task, information setting, general prompt instruction, and repeat number constant. Under this fixed comparison, the experiment contains seven generated implementations, one for each representation format.

We denote the hidden-test fraction for fixed comparison \(c\) and format \(f\) by \(s_{c,f}\). For each fixed comparison \(c\), let \(\bar{s}_c\) be the average hidden-test fraction across the seven formats in that comparison. The format effect for format \(f\) in fixed comparison \(c\) is \(s_{c,f} - \bar{s}_c\). A positive value means that the format scored above the seven-format average in that comparison, and a negative value means it scored below.

Some fixed comparisons do not distinguish the formats. If all seven formats obtain exactly the same hidden-test fraction in the same fixed comparison, then that comparison gives no information about their ordering. We therefore report two summaries. The first uses all fixed comparisons. The second is conditional on observing a nonzero format difference and uses only fixed comparisons in which at least two formats obtain different hidden-test fractions; we call these \emph{difference comparisons}. This second summary describes which formats score higher when the model is sensitive to format, and should not be read as an unconditional mean over all generations.

We also report pairwise win shares. For each pair of formats, we compare their hidden-test fractions within the same fixed comparison. Format \(f\) receives score \(1\) against format \(g\) if \(f\) has a higher hidden-test fraction, score \(1/2\) if the two formats tie, and score \(0\) otherwise. We average this score across fixed comparisons. In the tables, top share is the fractional share of fixed comparisons in which a format is tied for the highest hidden-test fraction. When multiple formats tie for the highest value, each tied format receives fractional credit. We compute confidence intervals by resampling fixed comparisons with replacement.

\subsection{Frequency of Format Differences}
\label{sec:frequency_format_differences}

Before comparing the seven formats, we check how often the formats lead to different hidden-test fractions within the same fixed comparison. Table~\ref{tab:informative} reports this count. For each model and information setting, there are 50 fixed comparisons because there are five tasks and 10 repeats.

When the prompt contains only the core algorithm description, format differences appear often. GPT-5.4 mini shows differences in 21 of the 50 fixed comparisons, Gemma 3 4B in 39, and Llama 3.2 3B in 38. When the prompt contains the complete specification, GPT-5.4 mini gives the same hidden-test fraction to all seven formats in all 50 fixed comparisons. In that setting, the format ordering among difference comparisons comes from Gemma 3 4B and Llama 3.2 3B.

\begin{table}[htbp]
\centering
\caption{Number of difference comparisons by model and information setting (Section~\ref{sec:comparison_unit}). The column ``With differences'' counts fixed comparisons in which at least two formats obtain different hidden-test fractions.}

\label{tab:informative}
\begin{tabular}{llrrr}
\toprule
Information setting & Model & Comparisons & With differences & Mean best-worst difference \\
\midrule
Algorithm-only & GPT-5.4 mini & 50 & 21 & 0.112 \\
Algorithm-only & Gemma 3 4B & 50 & 39 & 0.268 \\
Algorithm-only & Llama 3.2 3B & 50 & 38 & 0.250 \\
Complete & GPT-5.4 mini & 50 & 0 & 0.000 \\
Complete & Gemma 3 4B & 50 & 30 & 0.265 \\
Complete & Llama 3.2 3B & 50 & 38 & 0.317 \\
\bottomrule
\end{tabular}
\end{table}

\subsection{Core Algorithm Description}
\label{sec:core_algorithm_description}

The first main experiment uses prompts that contain the function name, Python signature, purpose, inputs, outputs, and computation steps. This setting measures the effect of the written form when the prompt gives the core algorithm description but does not add the extra implementation details used in the complete specification.

\begin{figure}[htbp]
\centering
\includegraphics[width=0.92\linewidth]{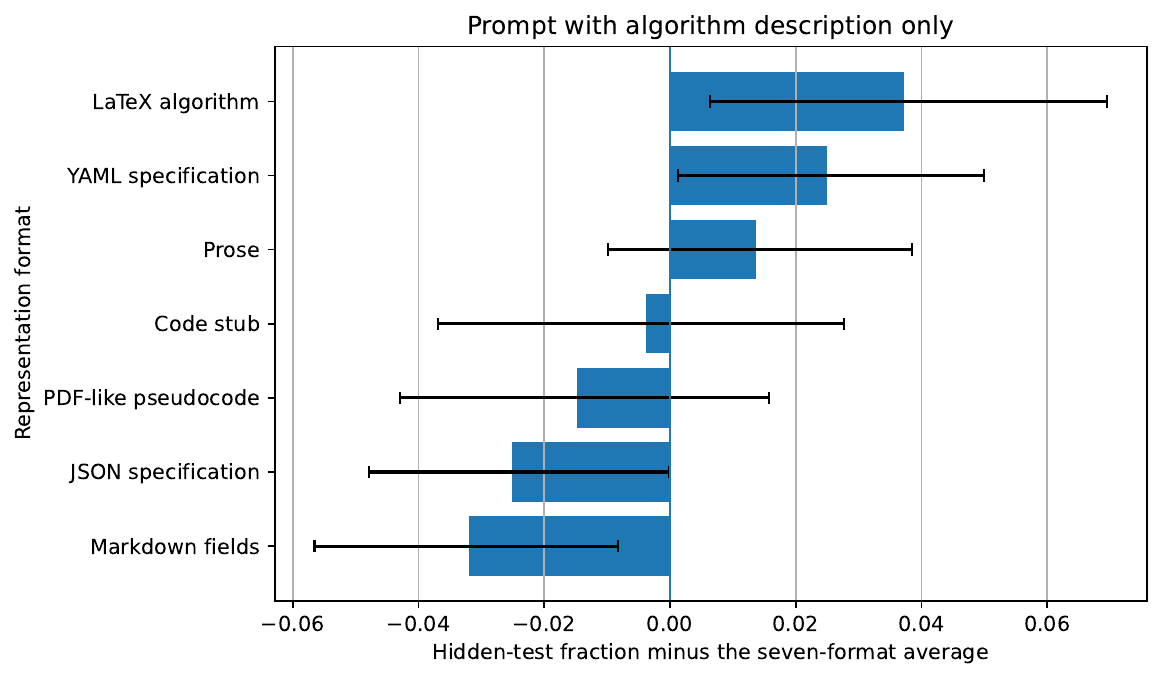}
\caption{Format effects when the prompt contains only the core algorithm description, computed over difference comparisons (Section~\ref{sec:comparison_unit}). Each value is the hidden-test fraction of the format minus the seven-format average in the same fixed comparison. Error bars are 95 percent intervals computed by resampling fixed comparisons.}
\label{fig:effect_algorithm_only}
\end{figure}

Figure~\ref{fig:effect_algorithm_only} and Table~\ref{tab:algorithm_only_leaderboard} summarize the difference comparisons. LaTeX algorithm-style pseudocode has the largest average format effect. Its mean hidden-test fraction is 0.384, and its average format effect is 0.037 with a 95 percent bootstrap interval from 0.010 to 0.069. YAML-like specification is second by average format effect, with mean hidden-test fraction 0.371 and average format effect 0.025. Ordinary prose is also competitive, with mean hidden-test fraction 0.360 and average format effect 0.014. Python code stub and PDF-like extracted pseudocode are close to the seven-format average in the same fixed comparisons. JSON-like specification and Markdown fields are below that average.

The strict all-hidden-tests-passed rate gives a different ordering. Code stub has the highest strict pass rate at 0.163, followed by YAML-like specification at 0.133, while LaTeX algorithm-style pseudocode has 0.102. Thus, the advantage of LaTeX algorithm-style pseudocode in this subsection is specific to the hidden-test fraction and to the average format effect computed within fixed comparisons. Under the stricter all-tests criterion, the ranking is different.

\begin{table}[htbp]
\centering
\caption{Summary for the core-algorithm-description experiment, computed over difference comparisons (Section~\ref{sec:comparison_unit}). Effect is the hidden-test fraction minus the seven-format average in the same fixed comparison. Top share is the fractional share of comparisons in which the format is tied for the highest hidden-test fraction.}
\label{tab:algorithm_only_leaderboard}
\begin{tabular}{lrrrr}
\toprule
Format & Hidden fraction & Hidden pass & Effect & Top share \\
\midrule
LaTeX algorithm & 0.384 & 0.102 & 0.037 & 0.211 \\
YAML specification & 0.371 & 0.133 & 0.025 & 0.147 \\
Prose & 0.360 & 0.112 & 0.014 & 0.152 \\
Code stub & 0.343 & 0.163 & -0.004 & 0.165 \\
PDF-like pseudocode & 0.332 & 0.112 & -0.015 & 0.139 \\
JSON specification & 0.321 & 0.092 & -0.025 & 0.093 \\
Markdown fields & 0.314 & 0.071 & -0.032 & 0.094 \\
\bottomrule
\end{tabular}
\end{table}

\begin{figure}[htbp]
\centering
\includegraphics[width=0.78\linewidth]{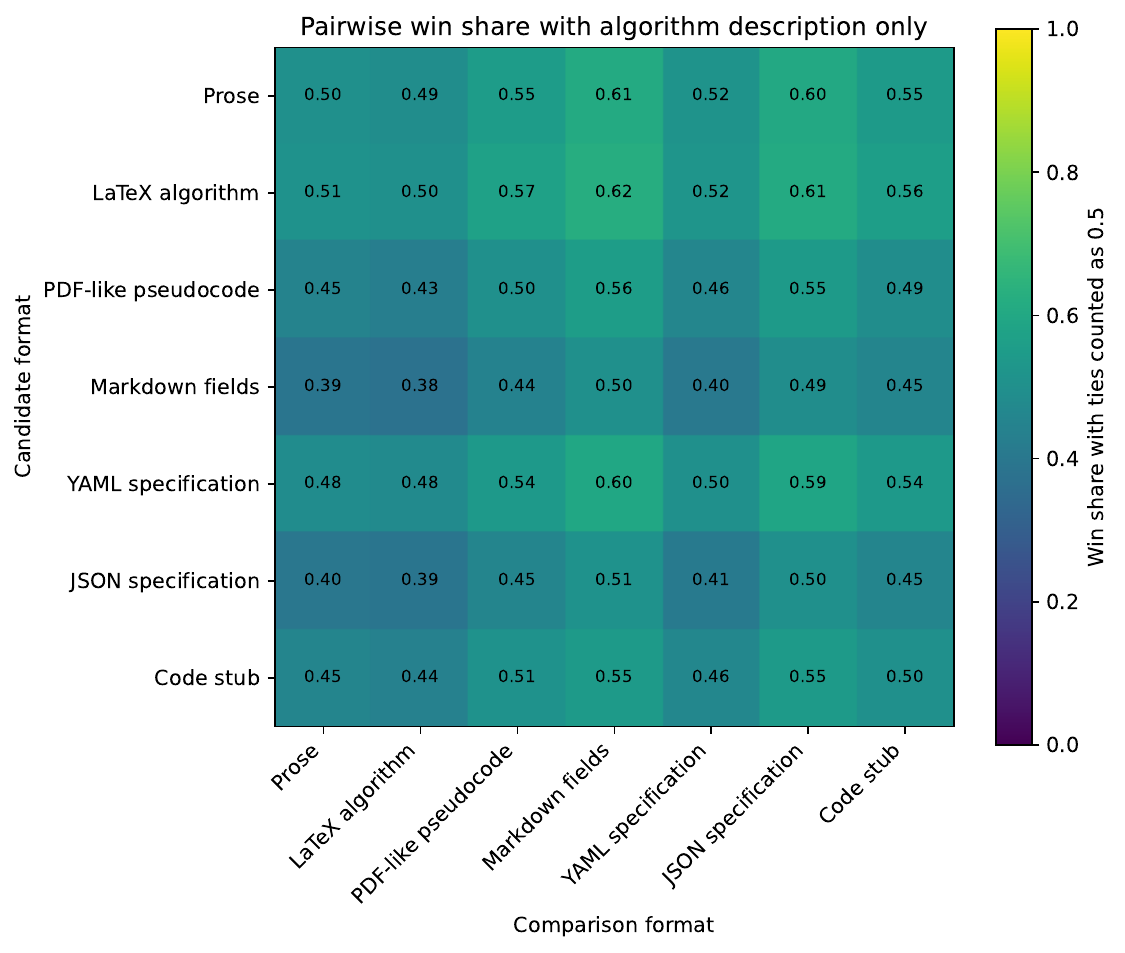}
\caption{Pairwise win share when the prompt contains only the core algorithm description, computed over difference comparisons (Section~\ref{sec:comparison_unit}). Rows are candidate formats and columns are comparison formats. Ties count as one half.}
\label{fig:pairwise_algorithm_only}
\end{figure}

The pairwise comparison in Figure~\ref{fig:pairwise_algorithm_only} gives a similar view. LaTeX algorithm-style pseudocode has an average pairwise win share of 0.566. Ordinary prose follows at 0.553, and YAML-like specification follows at 0.537. Markdown fields and JSON-like specification have the lowest average pairwise win shares. A pairwise win share of 0.5 means that two formats are tied on average after ties are counted as \(1/2\).

\subsection{Complete Specification}
\label{sec:complete_specification_results}

The second main experiment uses prompts that also contain configuration values, numerical rules, and invalid-input behavior. This setting asks whether format differences remain after implementation-relevant details are made more explicit.

\begin{figure}[htbp]
\centering
\includegraphics[width=0.92\linewidth]{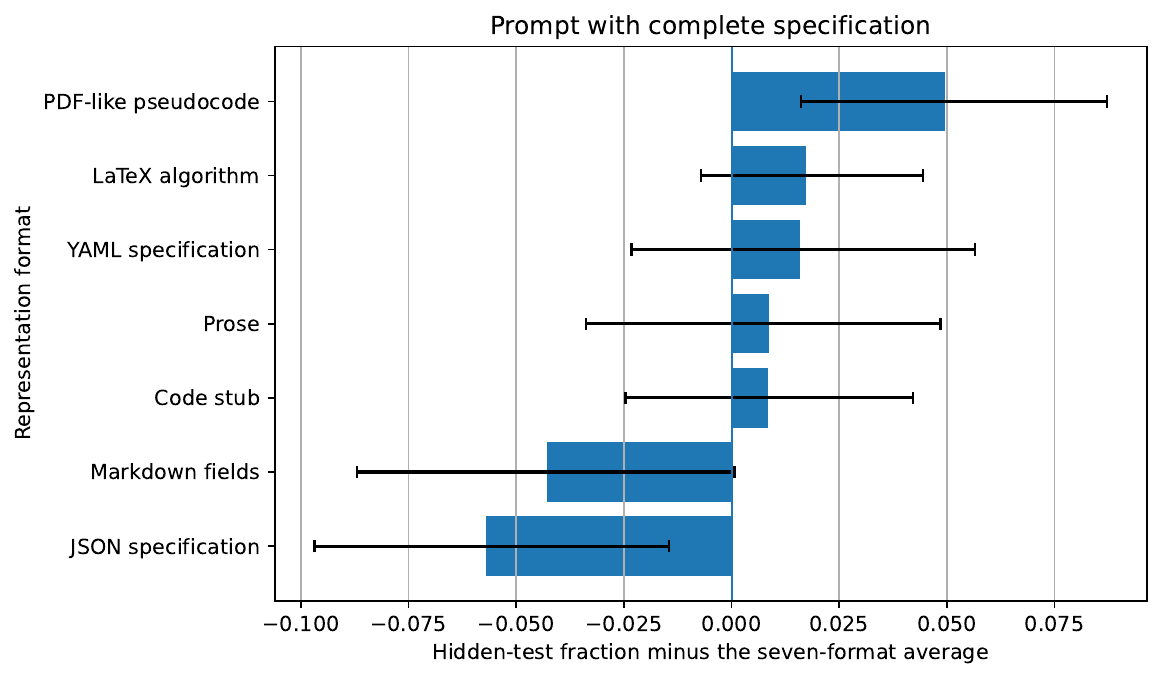}
\caption{Format effects when the prompt contains the complete specification, computed over difference comparisons (Section~\ref{sec:comparison_unit}). GPT-5.4 mini shows no format differences in this setting, so the figure reflects Gemma 3 4B and Llama 3.2 3B.}
\label{fig:effect_complete}
\end{figure}

Under complete information, GPT-5.4 mini gives the same hidden-test fraction to all seven formats in every task and repeat. The two smaller models behave differently: Gemma 3 4B shows differences in fewer fixed comparisons (30 rather than 39 of 50) with a similar mean best-worst gap, and for Llama 3.2 3B the frequency is unchanged (38 of 50) while the gap grows from 0.250 to 0.317 (Table~\ref{tab:informative}).
Figure~\ref{fig:effect_complete} and Table~\ref{tab:complete_leaderboard} therefore use only difference comparisons, which come entirely from these two models. 

Among those fixed comparisons, PDF-like extracted pseudocode has the largest average format effect at 0.049. LaTeX algorithm-style pseudocode, YAML-like specification, ordinary prose, and Python code stub are close to each other. Markdown fields and JSON-like specification are lower in this experiment.

\begin{table}[htbp]
\centering
\caption{Summary for the complete-specification experiment, computed over difference comparisons (Section~\ref{sec:comparison_unit}). GPT-5.4 mini shows no format differences in this setting, so the ordering is determined by Gemma 3 4B and Llama 3.2 3B.}
\label{tab:complete_leaderboard}
\begin{tabular}{lrrrr}
\toprule
Format & Hidden fraction & Hidden pass & Effect & Top share \\
\midrule
PDF-like pseudocode & 0.382 & 0.015 & 0.049 & 0.166 \\
LaTeX algorithm & 0.350 & 0.029 & 0.017 & 0.103 \\
YAML specification & 0.348 & 0.000 & 0.016 & 0.179 \\
Prose & 0.341 & 0.015 & 0.009 & 0.222 \\
Code stub & 0.341 & 0.029 & 0.009 & 0.120 \\
Markdown fields & 0.289 & 0.044 & -0.043 & 0.125 \\
JSON specification & 0.275 & 0.044 & -0.057 & 0.086 \\
\bottomrule
\end{tabular}
\end{table}

The PDF-like format here is a controlled linearization of the pseudocode block: the LaTeX environment is stripped and the steps are line-broken, but the text is not passed through a real parser or OCR pipeline. In Table~\ref{tab:complete_leaderboard}, PDF-like pseudocode leads by hidden-test fraction and average format effect among the difference comparisons. The strict pass metric gives a different view. Markdown fields and JSON-like specification have the highest all-tests-passed rate (0.044 each) in Table~\ref{tab:complete_leaderboard}, whereas PDF-like pseudocode, despite leading by hidden-test fraction, has a much lower rate (0.015), and YAML-like specification never passes all hidden tests in these comparisons (0.000) despite its high average fraction.

\subsection{Model-Specific Patterns}
\label{sec:model_specific_patterns}

The preceding summaries average over models whenever the fixed comparisons contain format differences. We therefore also examine which format scores highest within each model. This analysis shows whether the ordering of representation formats is stable across GPT-5.4 mini, Gemma 3 4B, and Llama 3.2 3B.

\begin{figure}[htbp]
\centering
\includegraphics[width=0.98\linewidth]{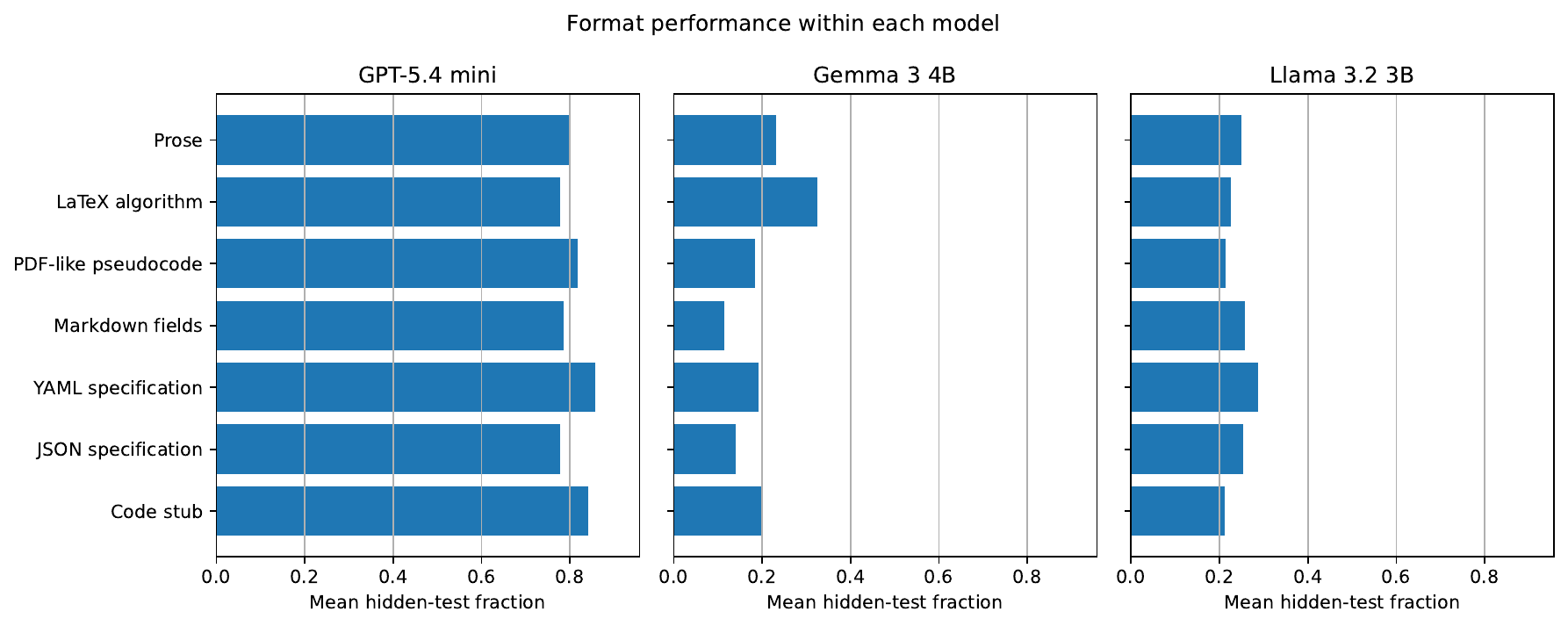}
\caption{Mean hidden-test fraction for each format within each model when the prompt contains only the core algorithm description. The order of formats is fixed across the three panels. The figure uses difference comparisons (Section~\ref{sec:comparison_unit}).}
\label{fig:model_algorithm_only}
\end{figure}

Figure~\ref{fig:model_algorithm_only} and Table~\ref{tab:model_top} show that the highest-scoring format differs across models. In the core-algorithm-description experiment, YAML-like specification is strongest for GPT-5.4 mini, LaTeX algorithm-style pseudocode is strongest for Gemma 3 4B, and YAML-like specification is strongest for Llama 3.2 3B, with Markdown fields close behind. In the complete-specification experiment, GPT-5.4 mini is not listed because all seven formats obtain the same hidden-test fraction in every fixed comparison. Among the remaining models, PDF-like pseudocode is strongest for Gemma 3 4B, while ordinary prose is strongest for Llama 3.2 3B. Because the top format changes across models, a single pooled ranking would hide these differences, so we report the per-model results alongside the pooled summary.

\begin{table}[htbp]
\caption{Top format for each model and information setting, computed over difference comparisons (Section~\ref{sec:comparison_unit}). GPT-5.4 mini is not listed for complete information because all seven formats obtain the same hidden-test fraction in every task and repeat.}

\centering
\small
\begin{tabular}{lllrr}
\toprule
Information setting & Model & Top format & Hidden fraction & Effect \\
\midrule
Algorithm-only & GPT-5.4 mini & YAML specification & 0.857 & 0.049 \\
Algorithm-only & Gemma 3 4B & LaTeX algorithm & 0.325 & 0.127 \\
Algorithm-only & Llama 3.2 3B & YAML specification & 0.287 & 0.044 \\
Complete & Gemma 3 4B & PDF-like pseudocode & 0.423 & 0.110 \\
Complete & Llama 3.2 3B & Prose & 0.396 & 0.048 \\
\bottomrule
\end{tabular}
\label{tab:model_top}
\end{table}

\subsection{Examples and Public Tests}
\label{sec:examples_public_tests}
The previous subsections compare representation formats under two information settings. We next examine how adding examples or public tests changes first-pass implementation accuracy under complete information.

The examples and public-test experiment uses k-means and attention. The examples setting gives the model small input-output examples in addition to the complete specification. The public-tests setting gives the model executable tests in addition to the complete specification. In both settings, the model generates code only once. It does not run the tests and does not repair the code from failure logs.

\begin{figure}[htbp]
\centering
\includegraphics[width=0.95\linewidth]{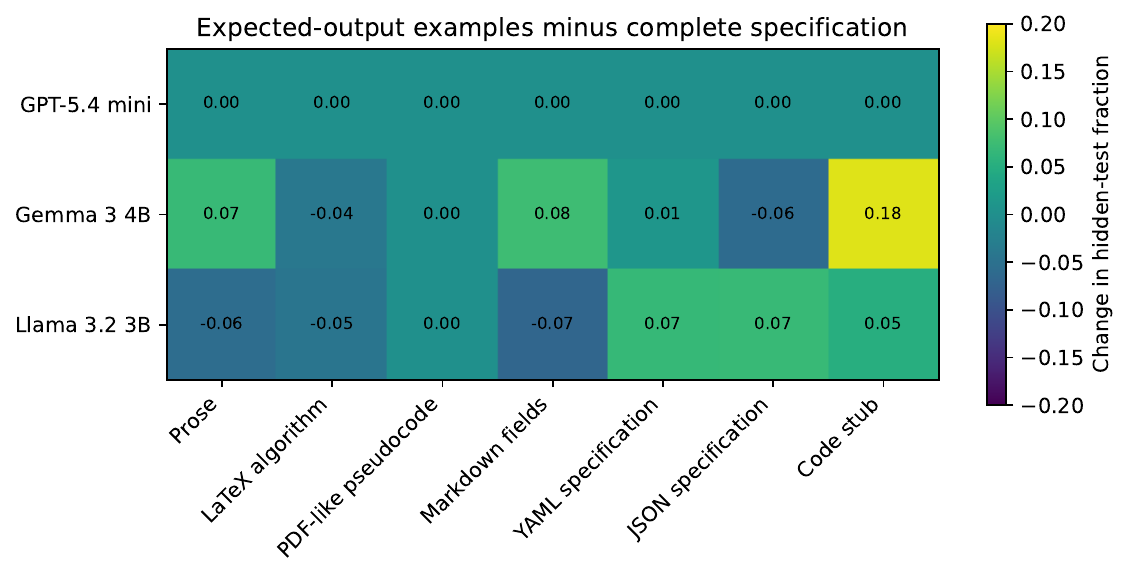}
\caption{Examples setting minus complete specification for k-means and attention. Values are changes in hidden-test fraction.}
\label{fig:examples_delta}
\end{figure}

\begin{figure}[htbp]
\centering
\includegraphics[width=0.95\linewidth]{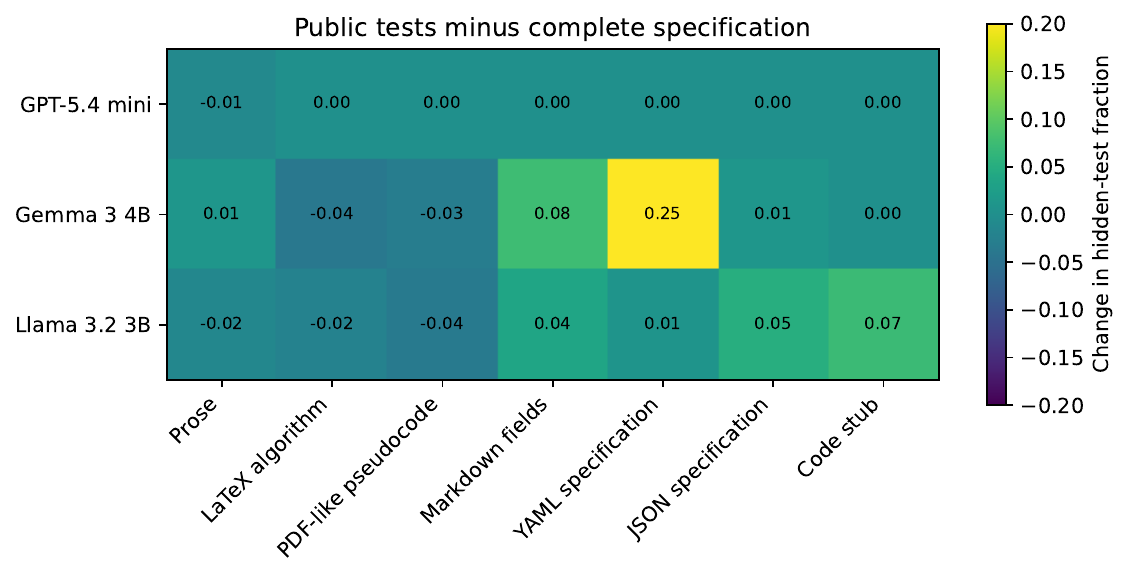}
\caption{Public-tests setting minus complete specification for k-means and attention. Values are changes in hidden-test fraction.}
\label{fig:public_tests_delta}
\end{figure}

For GPT-5.4 mini, expected-output examples leave the hidden-test fraction unchanged for all formats (Figure~\ref{fig:examples_delta}). Public tests change only ordinary prose, where the hidden-test fraction decreases by \(0.01\) (Figure~\ref{fig:public_tests_delta}). Under complete information, this model already gives identical scores across formats (Section~\ref{sec:complete_specification_results}). For Gemma 3 4B and Llama 3.2 3B, the changes are mixed. Public tests improve YAML-like specification for Gemma 3 4B by 0.250 hidden-test fraction relative to the complete specification, while they reduce PDF-like pseudocode for Llama 3.2 3B by 0.038. Examples improve Python code stub for Gemma 3 4B by 0.180, but reduce Markdown fields for Llama 3.2 3B by 0.068.

In this experiment, examples and public tests are used only as prompt text before the first generation. The model does not execute the public tests, observe failure logs, or revise the generated code. The mixed signs in Figures~\ref{fig:examples_delta} and~\ref{fig:public_tests_delta} therefore describe first-pass generation, while verification and repair with failure logs remain separate experimental settings.

\subsection{Field Removal}
\label{sec:field_removal}

The field-removal experiment evaluates YAML-like specification, JSON-like specification, and Python code stub for top-k distribution, k-means, and attention. Starting from the complete specification, the experiment removes one of four types of information: shapes, configuration values, numerical rules, or invalid-input behavior. The purpose is to examine whether these fields affect generated implementations, not to show that any field is safe to remove.

\begin{figure}[htbp]
\centering
\includegraphics[width=0.68\linewidth]{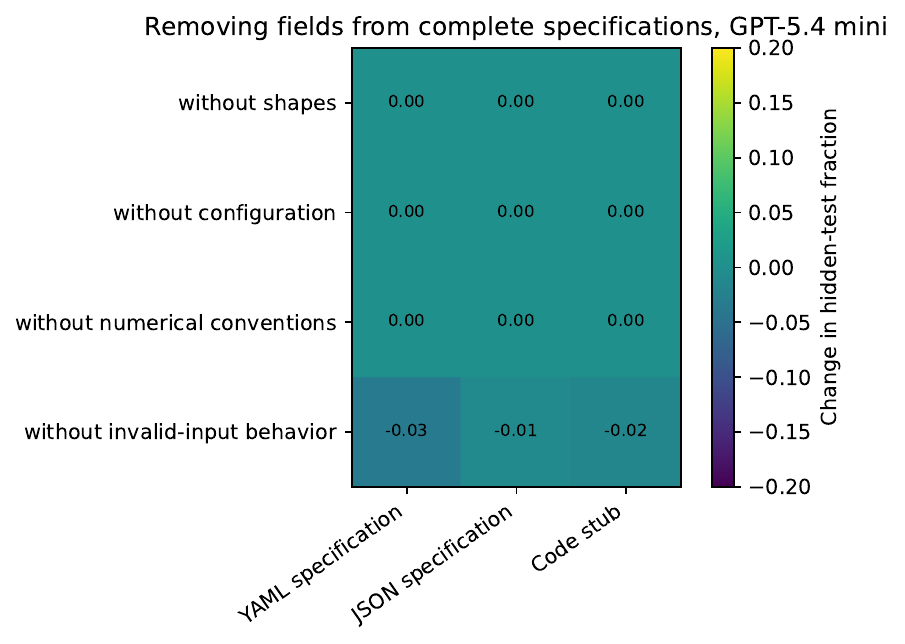}
\caption{Field-removal results for GPT-5.4 mini. Values are changes in hidden-test fraction relative to complete specifications for the same tasks and formats.}
\label{fig:field_ablation_gpt}
\end{figure}

For GPT-5.4 mini, removing invalid-input behavior has the clearest negative effect on the strict pass rate: the decrease is 0.067 for Python code stub, 0.033 for JSON-like specification, and 0.133 for YAML-like specification. The hidden-test-fraction changes in Figure~\ref{fig:field_ablation_gpt} are smaller, so this result mainly concerns complete success rather than the average fraction of hidden tests passed. For Gemma 3 4B and Llama 3.2 3B, the changes are not monotone. Some removals increase the average hidden-test fraction, so the field-removal experiment shows sensitivity to prompt composition rather than a monotone value of each field. Because the design changes both content and prompt length, it does not separate the effect of omitted information from the effect of shorter prompts.

\subsection{Prompt Length and Failure Diagnostics}
\label{sec:prompt_length_failure_diagnostics}

Finally, we check whether prompt length explains the preceding format differences and whether the hidden-test failures correspond to implementation-sensitive details.

Prompt length does not appear to account for the main ordering of formats. Within difference comparisons, the correlation between centered prompt length and centered hidden-test fraction is \(-0.083\) under core information and \(0.070\) under complete information. These correlations are small and have opposite signs across the two information settings.

\begin{table}[htbp]
\centering
\caption{Prompt-length diagnostics over difference comparisons (Section~\ref{sec:comparison_unit}). Scores and prompt lengths are centered by subtracting the average within the same fixed comparison.}
\label{tab:length}
\begin{tabular}{lrr}
\toprule
Information setting & Prompt-length correlation & Slope per 100 chars \\
\midrule
Algorithm-only & -0.083 & -0.010 \\
Complete & 0.070 & 0.010 \\
\bottomrule
\end{tabular}
\end{table}

Failure categories support the interpretation that hidden tests capture specification-sensitive details. Among failed generated implementations from the core-algorithm-description experiment, the most common hidden-test failure categories are invalid-input behavior, tie-breaking, and errors not assigned to a more specific category. In the complete-specification experiment, tie-breaking remains frequent, followed by errors not assigned to a more specific category and invalid-input behavior. Because the categories are derived from failed test names, they are rule-based diagnostics rather than manual annotations.

\section{Discussion and Writing Implications}
\label{sec:discussion_implications}
Taken together, the results support a deliberately narrow writing recommendation. Because format matters in many matched comparisons but its direction changes with the model and the information setting, no single surface format can be recommended on its own. Authors should therefore specify the choices that determine the implementation: interfaces, shapes, axes, numerical rules, tie-breaking, boundary cases, and invalid-input behavior.

\subsection{Interpretation of the Format Results}
\label{sec:format_interpretation}
In the core-algorithm-description experiment, LaTeX algorithm-style pseudocode has the largest average format effect, while YAML-like specification and ordinary prose remain close by hidden-test fraction. This pattern suggests that ordered computation steps and explicit inputs and outputs help even without a field schema. YAML-like specification performs strongly for GPT-5.4 mini and Llama 3.2 3B in the same setting, which supports the value of separating inputs, outputs, steps, and errors into named fields. The JSON-like results caution against attributing the YAML-like performance only to named fields. JSON-like specifications contain similar field labels but score lower here, so the experiment separates field structure from the surface syntax used to express it. The present design does not identify why JSON-like syntax performs worse.

The complete-specification experiment changes the interpretation. PDF-like pseudocode has the largest average format effect among difference comparisons, but those comparisons come from Gemma 3 4B and Llama 3.2 3B because GPT-5.4 mini shows no format differences in this setting. As noted in Section~\ref{sec:complete_specification_results}, this concerns linearized pseudocode under complete information, and does not measure the errors introduced by real PDF extraction.

A code stub fixes the interface, but its empty body does not state the computation rules that many hidden tests check, such as tie-breaking, numerical rules, and invalid-input behavior. The aggregate failure categories in Table~\ref{tab:failure_categories} show that these rules remain common sources of failure.

\subsection{Implications for Authors}
\label{sec:author_implications}
An implementation-oriented specification should state the function name and Python signature, then give the ordered computation steps together with the input shapes, output shapes, axes, and return structure. When they affect correctness, the specification should also state configuration values, tie-breaking rules, numerical rules, invalid-input behavior, and boundary cases. Examples and executable public tests are useful additions when reproducible implementation is part of the contribution.

These details need not appear in one block. The main text can use LaTeX algorithm-style pseudocode for readability, while supplementary material records function signatures, shapes, invalid-input behavior, examples, and public tests.

\subsection{Limitations and Future Work}
\label{sec:limitations_future_work}
This study uses function-level tasks. That choice makes the comparison controlled, but it does not cover the full difficulty of reproducing a large machine learning system from a paper. Accordingly, the conclusions concern algorithm specifications for well-defined functions rather than full-system research reproduction.

The study also evaluates a single generated implementation without repair. Public tests may become more useful when the model can see failure logs and revise the code. Finally, the PDF-like format is a controlled rendering of extracted pseudocode. Future experiments should test real PDF extraction and parsing pipelines, since those pipelines introduce their own errors.

\section{Conclusion}
This study shows that the written format of an algorithm specification can affect first-pass LLM implementation accuracy. In core-information prompts, LaTeX algorithm-style pseudocode has the largest average format effect, while YAML-like specifications and ordinary prose remain close in several comparisons. Under complete information, GPT-5.4 mini shows no format differences in the matched comparisons, but Gemma 3 4B and Llama 3.2 3B still do. Code stubs fix the signature but do not by themselves supply the computational rules needed for hidden-test success. The safest writing recommendation is therefore to state the function interface, shapes, axes, numerical rules, tie-breaking rules, and invalid-input behavior explicitly, whether these details appear in the main paper or in supplementary material.

\bibliography{arXiv2.bbl}
\bibliographystyle{tmlr}

\appendix

\section{Additional Tables}
The following tables provide additional diagnostics for hidden-test failures and the difference between public-test success and hidden-test failure.

\begin{table}[htbp]
\caption{Top hidden failure categories. Categories are derived from failed hidden-test names.}
\label{tab:failure_categories}
\centering
\small
\begin{tabular}{llrrr}
\toprule
Information setting & Category & Count & Failed implementations & Share \\
\midrule
Algorithm-only & Invalid-input behavior & 532 & 812 & 0.655 \\
Algorithm-only & Tie-breaking & 431 & 812 & 0.531 \\
Algorithm-only & Other & 275 & 812 & 0.339 \\
Algorithm-only & Configuration & 139 & 812 & 0.171 \\
Complete & Tie-breaking & 356 & 688 & 0.517 \\
Complete & Other & 249 & 688 & 0.362 \\
Complete & Invalid-input behavior & 206 & 688 & 0.299 \\
Complete & Configuration & 140 & 688 & 0.203 \\
\bottomrule
\end{tabular}

\end{table}

\begin{table}[htbp]
\centering
\caption{Public-to-hidden gap summaries for selected formats. The gap is the share of generated implementations in which public tests pass but hidden tests fail. The selected rows are included to illustrate the largest observed gaps and representative complete-information formats.}
\label{tab:public_hidden_gap}
\begin{tabular}{llrr}
\toprule
Information setting & Format & Public pass & Public-to-hidden gap \\
\midrule
Algorithm-only & Markdown fields & 0.500 & 0.300 \\
Algorithm-only & JSON specification & 0.487 & 0.273 \\
Algorithm-only & LaTeX algorithm & 0.493 & 0.273 \\
Algorithm-only & Prose & 0.493 & 0.267 \\
Complete & YAML specification & 0.580 & 0.247 \\
Complete & Prose & 0.553 & 0.213 \\
Complete & PDF-like pseudocode & 0.540 & 0.200 \\
Complete & Code stub & 0.493 & 0.153 \\
\bottomrule
\end{tabular}
\end{table}

\end{document}